# Simplified calcium signaling cascade for synaptic plasticity

Vladimir Kornijcuk[a], Dohun Kim[b], Guhyun Kim[b], Doo Seok Jeong[a]*

[c]Division of Materials Science and Engineering, Hanyang University, Wangsimni-ro 222, Seongdong-gu, 04763 Seoul, Republic of Korea

[b]School of Materials Science and Engineering, Seoul National University, Gwanak-no 1, Gwanak-gu, 08826 Seoul, Republic of Korea

*Corresponding author (D.S.J)
E-mail: dooseokj@hanyang.ac.kr

## Abstract

We propose a model for synaptic plasticity based on a calcium signaling cascade. The model simplifies the full signaling pathways from a calcium influx to the phosphorylation (potentiation) and dephosphorylation (depression) of glutamate receptors that are gated by fictive $C1$ and $C2$ catalysts, respectively. This model is based on tangible chemical reactions, including fictive catalysts, for long-term plasticity rather than the conceptual theories commonplace in various models, such as preset thresholds of calcium concentration. Our simplified model successfully reproduced the experimental synaptic plasticity induced by different protocols such as (i) a synchronous pairing protocol and (ii) correlated presynaptic and postsynaptic action potentials (APs). Further, the ocular dominance plasticity (or the experimental verification of the celebrated Bienenstock—Cooper—Munro theory) was reproduced by two model synapses that compete by means of back-propagating APs (bAPs). The key to this competition is synapse-specific bAPs with reference to bAP-boosting on the physiological grounds.

**Keywords**: synaptic plasticity, calcium signaling cascade, back-propagating action potential boost, synaptic competition

# 1. Introduction

Synaptic plasticity is the base of learning and endows the brain with high-level functionalities, such as perception, cognition, and memory.(Hebb, 1949; Kandel, 1978) The importance of synaptic plasticity has fueled vigorous investigations that have effectively revealed its nature from different perspectives over different subjects. In particular, alongside neuroscience, synaptic plasticity is also important as the basis for learning and inference in neuromorphic engineering (Mead, 1990) and computer science (artificial intelligence in particular). As such, rich theoretical models for synaptic plasticity have been proposed, including *macroscopic* models (Benuskova & Abraham, 2007; Clopath, Büsing, Vasilaki, & Gerstner, 2010; Froemke & Dan, 2002; Gjorgjieva, Clopath, Audet, & Pfister, 2011; Izhikevich & Desai, 2003; Pfister & Gerstner, 2006; Song, Miller, & Abbott, 2000; van Rossum, Bi, & Turrigiano, 2000) and *microscopic* models, such as a calcium signaling cascade models(Castellani, Bazzani, & Cooper, 2009; Cho, Aggleton, Brown, & Bashir, 2001; Graupner & Brunel, 2012; Hansel, Artola, & Singer, 1997; J Lisman, 1989; Shouval, Bear, & Cooper, 2002) and the computational neurogenetic modeling framework that connects neuronal and synaptic behaviors down to gene-level mechanisms(Benuskova & Kasabov, 2007; N. Kasabov, Benuskova, & Wysoski, 2005). The former class is phenomenological in that the microscopic pictures are often abstracted and recreated to formulate simple equation(s). Further, these macroscopic models can be categorized according to their description domain (activity or time domain). A typical measure of activity is the rate of action potentials (APs), i.e., the temporal average of spiking events in a particular time period. The activity domain models take neuronal activities as model variables. The seminal Hebbian learning (Cooper & Bear, 2012; Hebb, 1949) and modifications thereof, e.g., the Oja rule(Oja, 1982), Bienenstock-Cooper-Munro (BCM) rule (Bienenstock, Cooper, & Munro, 1982; Cooper & Bear, 2012), etc., belong to this class. A downside of these models is that the temporal structure of individual APs is ignored, resulting in conflict with temporal coding spiking neural networks (SNNs).(Jeong, 2018)

Explicit AP timing (rather than activity) matters in time domain models. Such models commonly involve kernels triggered by each AP to determine a change in synaptic weight upon the temporal distance between a presynaptic AP (preAP) and postsynaptic AP (postAP) pair.(Froemke & Dan, 2002;

Kistler & Hemmen, 2000; Song, et al., 2000) These are referred to as pair-based models for spike timing-dependent plasticity (STDP). The triplet model incorporates additional kernels to include preAP and postAP rates into the pair-based model.(Pfister & Gerstner, 2006) The purpose of this addition is to overcome the inability of simple pair-based models to reproduce physiological plasticity behaviors at high AP rates as analyzed by Izhikevich and Desai.(Izhikevich & Desai, 2003) Nevertheless, the kernels used in pair-based and triplet models lack physiological grounds.

Another class of neuron model and learning framework is the computational neurogenetic modeling framework that relates neuronal behaviors to gene-level mechanisms.(Benuskova & Kasabov, 2007; Nikola Kasabov, 2010; N. Kasabov, et al., 2005; N. K. Kasabov, 2019) The model is hierarchical in that the gene-level mechanisms underlie protein-level mechanisms that in turn underlie characteristic dynamics of membrane potential evolution. Therefore, this framework links the gene-level mechanisms with high-level brain functionalities (e.g., recognition) and macroscopic physiological brain behaviors. The framework includes various types of SNNs such as probabilistic SNN(Nikola Kasabov, 2010) and integrative probabilistic SNN(Benuskova & Kasabov, 2007). In spike of the finest domain ingredients in use in the framework, their behavioral abstraction reduces computational cost significantly.

Calcium-based models underpin their physiological grounds to some degree by incorporating calcium concentration into the synaptic plasticity dynamics.(Graupner & Brunel, 2012; Shouval, et al., 2002) These models consider the fact that the calcium concentration in the intracellular medium of a dendritic spine probably determines the synaptic plasticity because a high calcium concentration leads to long-term potentiation (LTP), whereas a low calcium concentration leads to long-term depression (LTD).(Cho, et al., 2001; Hansel, et al., 1997; J Lisman, 1989) A common strategy fixes the windows of calcium concentration for LTP and LTD, thereby changing the synaptic modification direction depending on the calcium concentration induced by preAPs and/or postAPs. Nevertheless, this simple physiological behavior cannot describe the rich dynamics of physiological synaptic plasticity, particularly the pathways from calcium to eventual synaptic plasticity, i.e., calcium signaling cascades.

Calcium signaling cascades have long been investigated to understand the direct origin of synaptic plasticity.(Castellani, et al., 2009; J Lisman, 1989; Lüscher & Malenka, 2012; Malenka & Nicoll, 1999; Nakano, Doi, Yoshimoto, & Doya, 2010) The proposed pathways initiated by a calcium influx involve a number of reversible and irreversible (primarily enzymatic)reactions in an attempt to understand the physiological cascades. As such, the dynamics involved are so rich that an intuitive understanding is considerably difficult. Furthermore, few calcium signaling cascade models consider the process preceding the cascades, i.e., a calcium influx by APs.(Castellani, et al., 2009; J Lisman, 1989; Lüscher & Malenka, 2012; Malenka & Nicoll, 1999; Nakano, et al., 2010) In this regard, two main issues arise in the simulation of large-scale SNNs. First, the complex implementation of calcium signaling cascades requires high-performance computation to decrease the gap between the physical time of the simulation and the computation runtime. Second, there is a lack of complete plasticity models described from a given preAP and postAP code (encoded as AP-timing and/or AP-rate) to its eventual synaptic plasticity through calcium signaling cascades.

To this end, we propose a complete excitatory synaptic plasticity model such that preAPs and postAPs are taken as inputs and eventually induce synaptic plasticity through simplified calcium signaling pathways. The pathways were simplified considering the following notable attributes in full pathways: competition between LTP and LTD over calcium ions and memory effect by irreversible enzyme reactions. Particularly, the second attribute results in a dynamic calcium concentration threshold for LTP, which is often fixed to a particular value in calcium-based synaptic plasticity models.(Graupner & Brunel, 2012; Shouval, et al., 2002) Additionally, the dynamic threshold offers a link with the BCM theory(Bienenstock, et al., 1982; Cooper & Bear, 2012) which incorporates a moving threshold for LTP. The proposed model was validated by applying three induction protocols on the physiological grounds. Note that our model takes protein-level ingredients in the finest domain unlike the framework of computational neurogenetic modeling. Also, we pay attention to the synaptic plasticity behavior (rather than high-level functionalities at the network scale) predicted by our model.

Section 2, Models, describes the proposed model in detail including the calcium influx mechanism and simplified calcium signaling cascades. Section 3, Results, is dedicated to simulated synaptic modification behaviors in response to the three induction protocols. Comparisons with previous models are elaborated in Section 4, Discussion, where additionally, the proposed model is discussed as an embedded learning algorithm in a neuromorphic hardware.

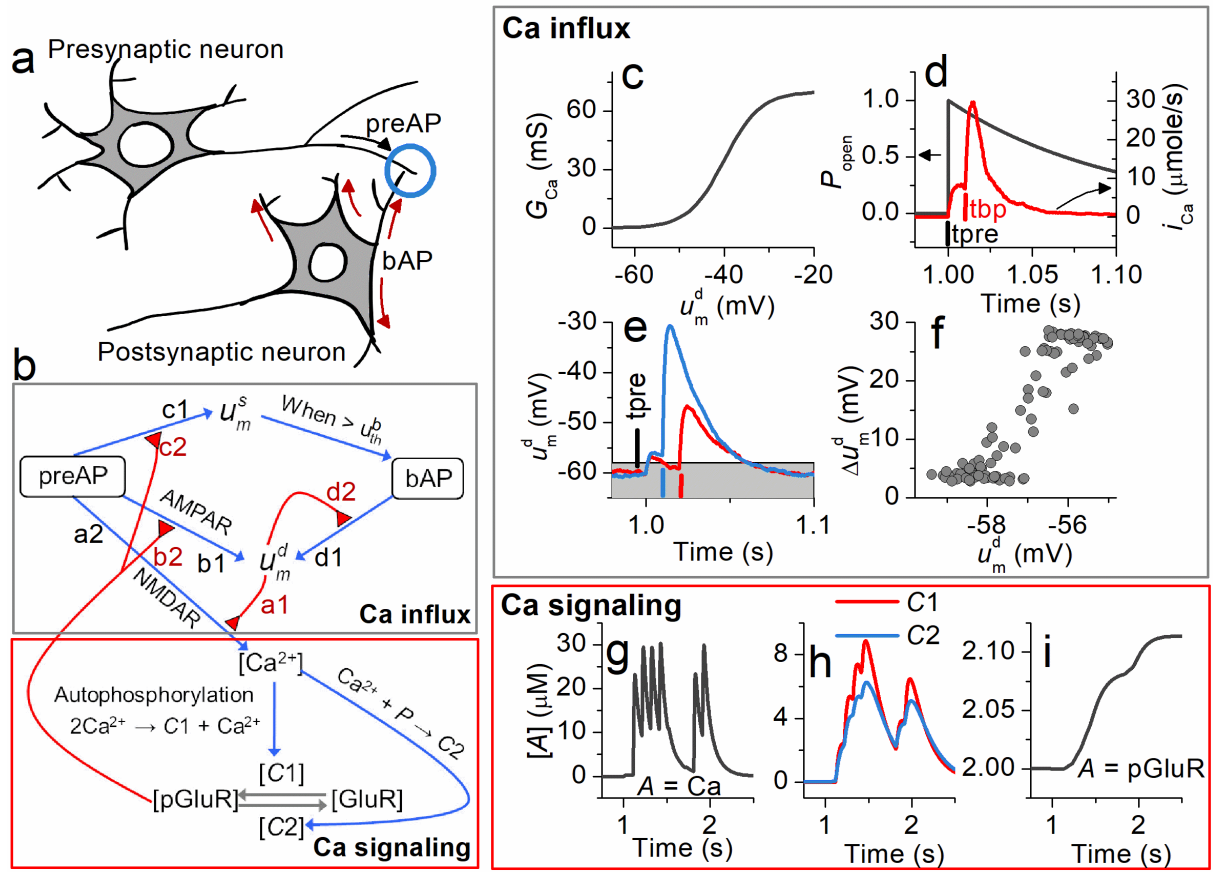

**Figure 1. Calcium influx and signaling pathway**. (**a**) Illustration of pre- and post-neurons. The blue circle indicates a synapse. (**b**) Schematic of calcium influx (upper panel) and signaling pathways (lower panel). The red triangles denote control gates. For instance, b2 indicates the somatic potential change gated by the pGluR concentration. (**c**) Conductance of calcium ions ($G_{Ca}$) over the active dendritic spine surface gated by the current dendritic potential. (**d**) (red curve) Calcium current change upon a preAP followed by a bAP (tbp – tpre = 10 ms), which is gated by linear filter $P_{open}$ (gray curve). (**e**) (blue curve) bAP boost above dendritic potential threshold (gray horizontal line at -58 mV) and boost failure below dendritic potential threshold (red curve). (**f**) Threshold dendritic potential for the bAP boost (indicated by a change in dendritic potential upon the arrival of a bAP). (**g**) Variation in [$Ca^{2+}$], (**h**) [$C1$] and [$C2$], and (**i**) [pGluR] with repeatedly applied preAP and postAP pairs ($\Delta t$ = tpost – tpre = 10 ms), which causes LTP.

## 2. Models

We used a multi-compartment (soma and dendritic spines) neuron model. All compartments of a neuron were described using a conductance-based model. Each dendritic spine has its own dendritic potential (cf. point neuron models). No dendritic APs are allowed because all APs are elicited from the soma. Such a somatic AP subsequently travels back to all relevant dendritic spines distributed along the dendrite, which is referred to as a back-propagating AP (bAP). The proposed synaptic plasticity model is spatially confined in the intracellular medium of a dendritic spine of a post-neuron, as shown in Fig. 1a. A calcium influx through the NMDAR channels and the consequent rise in calcium concentration in the dendritic spine trigger the proposed calcium signaling cascade, which eventually induces LTP or LTD. In this regard, the proposed model elucidates the calcium influx and subsequent calcium signaling mechanisms separately as schematized in the gray and red boxes in Fig. 1b, respectively. Hereafter, the molar concentration is expressed as [·]; e.g., the calcium concentration is expressed as [$Ca^{2+}$].

## 2.1. Neuron model

The model neuron consists of compartments (soma, dendritic spines, and axon). Given that one dendritic spine is assigned to each synapse, the more synapses the neuron holds, the more compartments are arranged in the neuron. The somatic potential $u_m^s$ is evaluated using a simple conductance-based neuron model (Fiete, Senn, Wang, & Hahnloser, 2010) with AP-rate adaptation as follows:

$$C_m^s \frac{du_m^s}{dt} = -g_l^s(u_m^s - u_l) - g_e^s(u_m^s - u_e) - g_{adp}^s(u_m^s - u_{adp}) + i_{noise}^s, \quad (1)$$

where $C_m^s$ is the capacitance of a somatic membrane. $g_l^s$ is the conductance of a base ionic leakage and $u_l$ is the equilibrium potential due to the base leakage only. $u_l$ is set to -65 mV so that the resting potential $u_{rest}$ is slightly above -65 mV because of the noise current $i_{noise}^s$ ($u_m^s \geq u_{rest} > u_l$). The second term on the right side of Eq. (1) implements the preAP-induced ionic leakage for all *n* synapses sharing the soma. All preAPs received by the *n* dendritic spines propagate to the soma, perturbing the somatic potential. Note that no delay in the propagation was assumed. $g_e^s$ denotes preAP-induced conductance of ionic leakage, and $u_e$ denotes the equilibrium potential due to this ionic leakage only. It holds that $u_e > u_m^s$ such that $u_e$ is zero, which is far above the depolarized potential. Therefore, this term causes the depolarization of the somatic membrane. The conductance $g_e^s$ is $g_e^s = g_{e0}^s \sum_{i=1}^{n} w_i s_i^s$, where $s_i^s$ is a kernel such that $s_i^s \leftarrow s_i^s + s_{i0}^s$ when a preAP is emitted and $ds_i^s/dt = -s_i^s/\tau_s$ otherwise. $g_{e0}^s$ is a constant and $w_i$ is the synaptic weight. The subscript $i$ in $s_i^s$ and $w_i$ denotes the synapse index.

The third term on the right side of Eq. (1) slows down the AP-induction rate. $g_{adp}^s$ is a kernel such that $g_{adp}^s \leftarrow g_{adp}^s + \Delta g_{adp}^s$ when the soma fires and $dg_{adp}^s/dt = -g_{adp}^s/\tau_{adp}$ otherwise. $u_{adp}$ is set below the somatic potential so that this adaptation term always inhibits the membrane depolarization, thereby reducing the AP-rate. The last term on the right side of Eq. (1) denotes a random electric current noise. The current was sampled every time bin from a Gaussian distribution centered at $\mu_i$ with $\sigma_i$ standard deviation. The soma emits an AP when the somatic potential reaches a spike threshold ($u_{th}$). The AP subsequently back-propagates to all dendritic spines, which consequently perturbs their dendritic potentials. For simplicity, no delay in back-propagation is assumed. A refractory period ($t_{ref}$) is immediately given to the soma.

The potential of the $i^{th}$ dendritic spine $u_{m,i}^d$ was acquired from the equation

$$C_m^d \frac{du_{m,i}^d}{dt} = -g_l^d(u_{m,i}^d - u_l) - g_{e,i}^d(u_{m,i}^d - u_e) + i_{noise}^d,$$

where the first term on the right side is the base ionic leakage, as for the somatic potential in Eq. (1); the second term implements the preAP-induced ionic leakage. The conductance $g_{e,i}^d$ is determined by preAPs and bAPs as follows

$$g_{e,i}^d = g_{e0}^d(w_i \cdot s_i^d + s_i^b),$$

where $g_{e0}^d$ and $s_{i0}^d$ are constants. $s_i^d$ is a kernel such that $s_i^d \leftarrow s_i^d + s_{i0}^d$ when a preAP is emitted from the pre-neuron, and $ds_i^d/dt = -s_i^d/\tau_d$ otherwise. bAPs are integrated by $s_i^b$ such that $s_i^b \leftarrow s_i^b + \Delta s_i^b$ upon the arrival of a BP_AP and $ds_i^b/dt = -s_i^b/\tau_{bp}$ otherwise, where $\Delta s_i^b$ is a function of the dendritic potential as

$$\Delta s_i^b = s_0^b / \left[1 + e^{-k(u_{m,i}^d - u_{th}^b)}\right] + s_1^b.$$

Here, $s_0^b$, $k$, and $s_1^b$ are constants (>0) and $s_0^b > s_1^b$. $u_{th}^b$ is a threshold dendritic potential for a bAP boost. If the dendritic potential is larger than the threshold, $\Delta s_i^b$ is also large and the bAP significantly increases the dendritic potential. Otherwise, the contribution of the bAP to the dendritic potential is smaller because $s_0^b > s_1^b$. Analogous to the somatic potential, electric current noise is implemented by $i_{noise}^d$ [~$N(\mu_i, \sigma_i)$].

The model parameters are listed in Table 1. Note that the $C_m$, $s_{i0}$, and $\tau$ values for the dendritic potential evaluation were different from the somatic potential evaluation to consider the difference in excitatory postsynaptic potential (EPSP) evolution between the dendrite and soma.(Spruston, 2008)

## 2.2. Calcium influx through NMDAR channels

On the grounds of physiological evidence, the permeability (conductance) of the NMDAR channels to calcium ions is raised by the release of neurotransmitters in conjunction with a simultaneous rise in dendritic potential $u_m^d$, referred to as dendritic coincidence detection.(G. J. Stuart & Hausser, 2001) In our model, the conductance $g_{Ca}$ is given by a multiplicative function of $G_{Ca}$ (only dependent on the dendritic potential) and kernel $P_{open}$, as $g_{Ca} = G_{Ca} P_{open}$. To maintain physiological plausibility(Jahr & Stevens, 1990), $G_{Ca}$ was set to a logistic function of dendritic potential with a maximum of $G_{Ca0}$ as

$$G_{Ca} = G_{Ca0} / \left[1 + e^{-k_{Ca}(u_m^d - u_{Ca})}\right], \quad (2)$$

where $k_{Ca}$ and $u_{Ca}$ are constants. This function is plotted in Fig. 1c. Its effect on [Ca$^{2+}$] is indicated by a1 in Fig. 1b. The kernel $P_{open}$ defines the temporal distance and order between a preAP and bAP such that it reaches its maximum (= 1) when a preAP is elicited and decays exponentially with a time constant of $\tau_{open}$. This effect of a preAP on the calcium influx and resulting change in [Ca$^{2+}$] are indicated by a2 in Fig. 1b and the behavior of the kernel is shown in Fig. 1d. Given the multiplicative relationship, the process a2 is gated by the dendritic potential through $G_{Ca}$ (a1 in Fig. 1b).

Finally, the calcium current through NMDAR channels ($i_{Ca}$) was assumed to conform to Ohm's law as

$$i_{Ca} = g_{Ca}(u_m^d - u_m^{d0}), \qquad (3)$$

where $u_m^{d0}$ denotes the dendritic potential in equilibrium between a calcium influx and outflux. For instance, Fig. 1d shows a time-dependent profile of the calcium current driven by a preAP and bAP pair separated by 10 ms ($tbp - tpre = 10\ ms$), where tbp and tpre denote the bAP and preAP timing, respectively.

The direct causes of dendritic potential change are the preAP and bAP. A preAP causes an excitatory postsynaptic current (EPSC) through the AMPAR channels and the consequent evolution of a dendritic EPSP, indicated by b1 in Fig. 1b. The dendritic EPSP depends on the synaptic weight, which is proportional to the concentration of phosphorylated glutamate receptors (pGluR), gating the b1 process through b2 in Fig. 1b.

Both the dendritic and somatic potentials are raised by preAPs, as indicated by c1 in Fig. 1b. Likewise, the effect is gated by the synaptic weight, which is a function of [pGluR], as denoted by c2. However, their effect on the somatic EPSP is assumed to be smaller than that on the dendritic EPSP based on the potential attenuation along the dendrite toward the soma.(G. Stuart & Spruston, 1998) The somatic EPSP by each preAP undergoes temporal integration until the resulting value exceeds a threshold for spiking and a somatic AP is generated. The bAP is the other primary cause of change in dendritic potential because it depolarizes the dendritic spine (indicated by d1 in Fig. 1b). Note that the model ignores delays in bAP so that a bAP immediately arrives at its target dendritic spines after postAP generation.

The bAP perturbs the dendritic potentials to different extents such that only dendritic potential over a certain threshold ($u_{th}^b$) enables a bAP boost(Sjöström & Häusser, 2006; Sjöström, Turrigiano, &

Nelson, 2001) causing a large increase in the dendritic potential. An example is plotted in Fig. 1e, where each of the two dendritic potentials initially perturbed by simultaneous preAPs at tpre (1 s) encounters a bAP at a different tpost (1.01 and 1.02 s for the blue and red curves, respectively) at which the dendritic potential is also different (the blue curve remains above a bAP boost threshold of -58 mV, whereas the red one falls below the threshold). The changes in dendritic potential upon bAP generation ($\Delta u_m^d$), which is a measure of the bAP boost, for different dendritic potentials are summarized in Fig. 1f. The variability in the data is due to the stochasticity of the threshold value (conforming to a Gaussian distribution with -58 mV mean and 5.8 mV standard deviation). Therefore, the dendritic potential controls the effect of bAP on itself, as indicated by d2 in Fig. 1b. The implication is that the d1 and d2 processes set a positive feedback loop through a highly depolarized dendritic spine ($u_m^d > u_{th}^b$), which consequently gains more synaptic weight due to a significant influx of calcium ions (a1 in Fig. 1b). By contrast, a dendritic spine with insufficient depolarization for a bAP boost ($u_m^d < u_{th}^b$), likely implying its minor contribution to the bAP generation, loses weight by the bAP, causing a small influx of calcium ions and the consequent decrease in weight. Therefore, the bAP boost dependence on the dendritic potential sets competition between dendritic spines sharing the same soma. This is the key to the reproduction of the monocular deprivation behavior(Mioche & Singer, 1989), which will be discussed in detail in the Discussion section.

Alongside the random variability in the bAP boost threshold, stochasticity in dendritic and somatic potentials is present such that a random current is applied to both dendritic and somatic compartments, which follows a Gaussian distribution centered at 9 mA with standard deviations of 18 and 72 mA for dendritic and somatic potentials, respectively.

## 2.3. Calcium signaling cascade

When depolarized, the dendritic spine allows a calcium influx through NMDAR channels, which eventually leads to long-term plasticity through calcium signaling pathways.(John Lisman, Yasuda, & Raghavachari, 2012; Nakano, et al., 2010; Oliveira, Kim, & Blackwell, 2012; Pepke, Kinzer-Ursem,

Mihalas, & Kennedy, 2010) We simplified such pathways in terms of the following features of biological signaling cascades (the red box in Fig. 1b)

**Autophosphorylation of CaMKII**. A calcium influx is a trigger for calcium-calmodulin-dependent protein kinase II (CaMKII) activation, which subsequently activates its autophosphorylation. This is the direct cause of phosphorylation of AMPA glutamate receptors (GluRs).(John Lisman, et al., 2012; Lüscher & Malenka, 2012; Nakano, et al., 2010; Oliveira, et al., 2012)

**Enzymatic reaction of PP and $Ca^{2+}$ or $Ca^{2+}$/CaM**. Calcium injection is also a trigger for protein phosphatases (PPs), such as calcineurin (CaN), activation through either its direct binding to PP or its binding to CaM and subsequent binding of $Ca^{2+}$/CaM to PP.(Baumgärtel & Mansuy, 2012; Beattie, et al., 2000; Claude B. Klee, Ren, & Wang, 1998; Lüscher & Malenka, 2012; Nakano, et al., 2010; Stemmer & Klee, 1994) The activated PP dephosphorylates GluRs (depression).

**Phosphorylation and dephosphorylation of GluRs (memory effect)**. A reaction of phosphorylated CaMKII (pCaMKII: enzyme) with GluRs (substrate) to phosphorylate GluR (resulting in pGluR) gives rise to a memory effect due to the irreversible reaction from the complex to the product. The same holds for the dephosphorylation of pGluRs by PP (resulting in GluR). Thus, pCaMKII and PP compete for GluRs.

Our model includes $Ca^{2+}$, GluR, pGluR, fictive catalysts $C$1 and $C$2, and the fictive element $P$.

$$\begin{cases} Ca^{2+} \xrightarrow{PR1} Ca^{2+}/CaM \xrightarrow{PR2} active\ CaMKII \xrightarrow{PR3} pCaMKII \\ \qquad\qquad Ca^{2+} \xrightarrow{SPR1} C1 \end{cases} \text{for LTP} \qquad (4)$$

Equation (4) is a representative path for LTP.(Bradshaw, Kubota, Meyer, & Schulman, 2003; John Lisman, et al., 2012; Lüscher & Malenka, 2012; Nakano, et al., 2010; Pepke, et al., 2010) The reaction PR3 in Eq. (4) denotes the autophosphorylation of active CaMKII, which can be modeled as an enzymatic reaction between two active CaMKII species.(Hanson, Meyer, Stryer, & Schulman, 1994) This pathway is simplified to the lower equation in Eq. (4), in which calcium ions are directly converted to catalyst $C$1 (equivalent to pCaMKII) as a result of the reaction SPR1. The reaction SPR1 captures the autophosphorylation (i.e., autocatalysis) of active CaMKII such that

$$2Ca^{2+} \to C1 + Ca^{2+}. \tag{5}$$

In turn, fictive catalyst $C$1 phosphorylates GluR,

$$GluR + C1 \to pGluR + C1, \tag{6}$$

leading to potentiation. The irreversible reaction in Eq. (6) highlights the memory effect of [pGluR] given that, when [$C$1] vanishes, the induced [pGluR] is maintained, unlike the reversible chemical reaction embodying paired forward and reverse reactions. This is the direct cause of LTP.

$$\begin{cases} Ca^{2+} \to Ca^{2+}/CaM \to active\ CaN \\ \quad Ca^{2+} \xrightarrow{SDR1} C2 \end{cases} \quad \text{for LTD}, \tag{7}$$

CaN is the key PP for LTD in calcium signaling.(Baumgärtel & Mansuy, 2012; Beattie, et al., 2000; Lüscher & Malenka, 2012; Nakano, et al., 2010; Stemmer & Klee, 1994) The upper equation in Eq. (7) indicates the activation of CaN through $Ca^{2+}$/CaM, which gates the affinity of CaN for calcium given the interactive change in the conformations of the subunits.(C B Klee, Crouch, & Krinks, 1979; Claude B. Klee, et al., 1998; Stemmer & Klee, 1994) Likewise, active CaN is replaced by $C$2, as shown in Eq. (7), and the reaction SDR1 is given by the following simple enzymatic reaction,

$$Ca^{2+} + P \to C2, \tag{8}$$

where $P$ denotes the fictive protein phosphate directly involved in the simplified reaction that leads to LTD.

The consequent $C$2 dephosphorylates pGluR as

$$pGluR + C2 \to GluR + C2, \tag{9}$$

leading to depression. This one-way reaction also causes the memory effect, leading to LTD.

Figure 1g shows a calculated response of [Ca$^{2+}$] to ten preAP and bAP pairs ($\Delta t$ = tbp – tpre = 10 ms) repeatedly applied at 10 Hz to a synapse. Given no delay in AP back-propagation, tbp equals the postAP timing. The causal order ($\Delta t > 0$) and small $\Delta t$ between the preAP and bAP satisfies the dendritic coincidence condition, leading to a high calcium influx. The consequent changes in [$C1$] and [$C2$] through the proposed calcium signaling pathways (reactions (5) and (8)) confirm that $C1$ outperforms $C2$ (Fig. 1h). Thus, an increase in [pGluR] by reaction (6) outweighs the opposite reaction (9), thereby leading to LTP (Fig. 1i).

## 2.4. Formulation

The time-dependent change in [Ca$^{2+}$] in the dendritic spine can be expressed as

$$\frac{d[Ca^{2+}]}{dt} = \frac{i_{Ca}}{V_{sp}} - \frac{[Ca^{2+}]}{\tau_{out(Ca)}} - k_{p1}^f[Ca^{2+}]^2 - k_{d1}^f[Ca^{2+}][P], \tag{10}$$

where $\tau_{out(Ca)}$ denotes an escape time constant of calcium ions from the dendritic spine (whose volume is $V_{sp}$) to the parent dendrite. The first term of the right side of Eq. (10) indicates a calcium source due to the calcium influx by Eqs. (2) and (3) and the other terms indicate calcium sinks. Calcium ions in the spine are uncaged in due course with a time constant of $\tau_{out(Ca)}$, which is indicated by the second term. Furthermore, the proposed calcium signaling cascade consumes calcium ions in the spine by reactions (5) and (8). Those effects are quantified by the third and fourth terms, respectively. Therefore, $k_{p1}^f$ and $k_{d1}^f$ denote the rate constants of reactions (5) and (8), respectively.

[$C1$] and [$C2$] are given by

$$\frac{d[C1]}{dt} = -\frac{[C1]}{\tau_{out(C1)}} + k_{p1}^f[Ca^{2+}]^2 \tag{11}$$

and

$$\frac{d[C2]}{dt} = -\frac{[C2]}{\tau_{out(C2)}} + k_{d1}^f [Ca^{2+}][P] , \tag{12}$$

respectively. $\tau_{out(C1)}$ and $\tau_{out(C2)}$ are the escape time constants of the catalysts $C1$ and $C2$, respectively. Finally, from reactions (6) and (9), [pGluR] is expressed as

$$\frac{d[pGluR]}{dt} = k_{p2}^f [C1]([GluR]_0 - [pGluR]) - k_{d2}^f [C2][pGluR] , \tag{13}$$

where $k_{p2}^f$ and $k_{d2}^f$ are the rate constants of reactions (6) and (9), respectively. We limit the total concentration of GluRs subject to phosphorylation to $[GluR]_0$. Thus, the available [GluR] for further phosphorylation is $[GluR]_0 - [pGluR]$.

The model parameters used in this study are listed in Table 1.

## 3. Results

The proposed model was subject to three synaptic modification protocols. Protocol 1 is a standard STDP induction method, where the difference between preAP and postAP timings is controlled precisely.(Bi & Poo, 1998; Froemke & Dan, 2002; Markram, Lübke, Frotscher, & Sakmann, 1997; Sjöström, et al., 2001) In this case, the postAP timing equals the bAP timing because there is no delay in AP back-propagation. Protocol 1 considers the rate of the preAP and postAP pair as an additional variable to identify the intermingling of the timing-difference and AP-rate effects.(Sjöström, et al., 2001) Thus, it rules out any irregularity in spiking behavior such as random variability in interspike intervals (ISIs), leaving the temporal correlation between a preAP and postAP under strict control. Protocol 2 takes the preAP rate (with perfect periodicity) as the only variable while leaving the spiking behavior of the post-neuron to be determined by the preAPs and noise current. That is, unlike Protocol 1, the preAP-postAP temporal correlation is not directly controlled by the protocol; instead, the correlation is naturally established (if possible) by the postAPs induced by the periodic preAPs. In this regard, this protocol situates the synaptic plasticity model in more natural circumstances compared with Protocol 1. Protocol 2 has often been applied in an attempt to induce a firing-rate-dependent crossover between homosynaptic LTD and LTP as clear evidence for the BCM theory.(Dudek & Bear, 1992; Philpot, Cho, & Bear, 2007)

While Protocols 1 and 2 address unitary synaptic connection (between a single pre- and post-neuron), Protocol 3 addresses the case where multiple pre-neurons share a single post-neuron. For simplicity, we assume two pre-neurons (each of which is given one synapse) and one post-neuron. Each of the pre-neurons fires Poisson APs (rather than periodic ones) with a given rate and postAPs are induced because of the preAPs and noise current. In a broad sense, the application domain of Protocol 3 includes the ocular dominance plasticity experiment.(Mioche & Singer, 1989)

The three protocols are summarized in Table 2. Following each synaptic plasticity process, the plasticity was evaluated by measuring the deviation of the maximum somatic EPSP from the initial one in response to a single read-out preAP. Hereafter, the plasticity behaviors induced by Protocols 1, 2, and 3 are referred to as Behaviors 1, 2, and 3, respectively.

The simulated Behaviors 1 and 2 are compared with experimental data to justify the proposed plasticity model. We address the plasticity behavior of cortical neurons in response to Protocol 1, reported by Sjöström, Turrigiano, and Nelson(Sjöström, et al., 2001). To justify Behavior 2, we address the homosynaptic plasticity of hippocampal neurons, observed by Dudek and Bear.(Dudek & Bear, 1992) It should be noted that the similarity of the simulated behaviors to physiological data considers general behavioral tendencies rather than specific behaviors. This is because the same specific behavior is rarely considered in more than one publication. However, several common tendencies in the observations hold.

Regrettably, real-time monitoring data of the weight evolution in response to Protocol 3 is unavailable. Instead, Behavior 3 is verified with reference to the simulation results of plasticity using the BCM theory(Cooper & Bear, 2012), which successfully reproduces physiological ocular dominance data, such as normal rearing, monocular deprivation, reverse suture, and binocular deprivation.

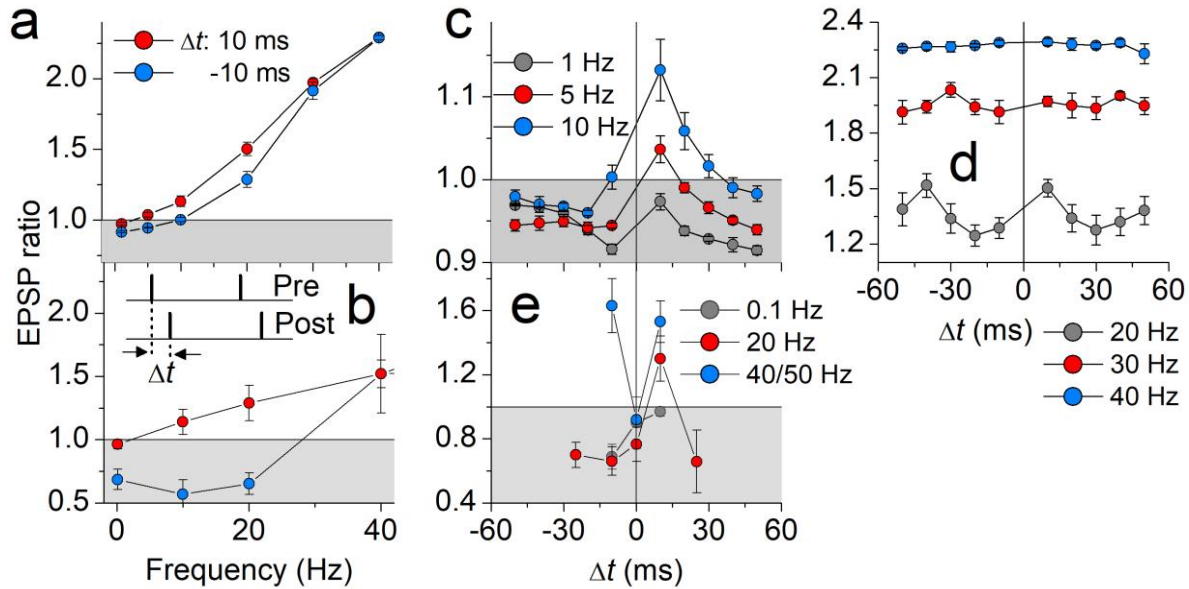

**Figure 2. Long-term plasticity induced by Protocol 1**. (**a**) Long-term plasticity induced by 40 repeatedly applied preAP and postAP pairs, each of which was separated by 10 ms (red-filled circles) or -10 ms (blue-filled circles). The applied frequency varied from 1 to 40 Hz. (**b**) Experimental plasticity behavior of cortical neurons in response to Protocol 1. A temporal configuration of a preAP and postAP is schematized in the inset. The long-term plasticity induced by different spike timings (-50–50 ms) at (**c**) 1, 5, 10, (**d**) 20, 30, and 40 Hz are also plotted. For comparison, the plasticity behaviors of cortical neurons at 0.1, 20, 40, and 50 Hz are shown in (**e**). The data in (**b**) and (**e**) are taken from Sjöström, Turrigiano, and Nelson (Sjöström, Turrigiano, & Nelson, 2001).

### 3.1. Protocol 1

The temporal order and distance between a preAP and postAP (thus bAP), parameterized by $\Delta t$ (= tpost – tpre), were controlled externally, i.e., the temporal correlation between the two events was predefined. We varied $\Delta t$ in the range between -50 and 50 ms) and, for all the synaptic plasticity induction runs, we applied 40 preAP and postAP pairs. Additionally, the rate (frequency) of these preAP and postAP pairs was varied from 1 to 40 Hz. The ratio of EPSP maximum after plasticity to initial EPSP maximum was taken as a plasticity indicator because ratios below and above unity indicate LTD and LTP, respectively. The simulation results for two $\Delta t$ values (-10 and 10 ms) are plotted in Fig. 2a. LTD is obvious at frequencies below 10 Hz and $\Delta t$= -10 ms. In other words, LTD vanishes when the frequency exceeds 10 Hz irrespective of the temporal order between a preAP and postAP (-10 ms or 10 ms), and thus only LTP is induced. This is akin to the physiological observations by Sjöström, Turrigiano, and Nelson(Sjöström, et al., 2001) as plotted in Fig. 2b. The data identify vanishing LTD for $\Delta t$ = -10 ms at a frequency of 40 Hz.

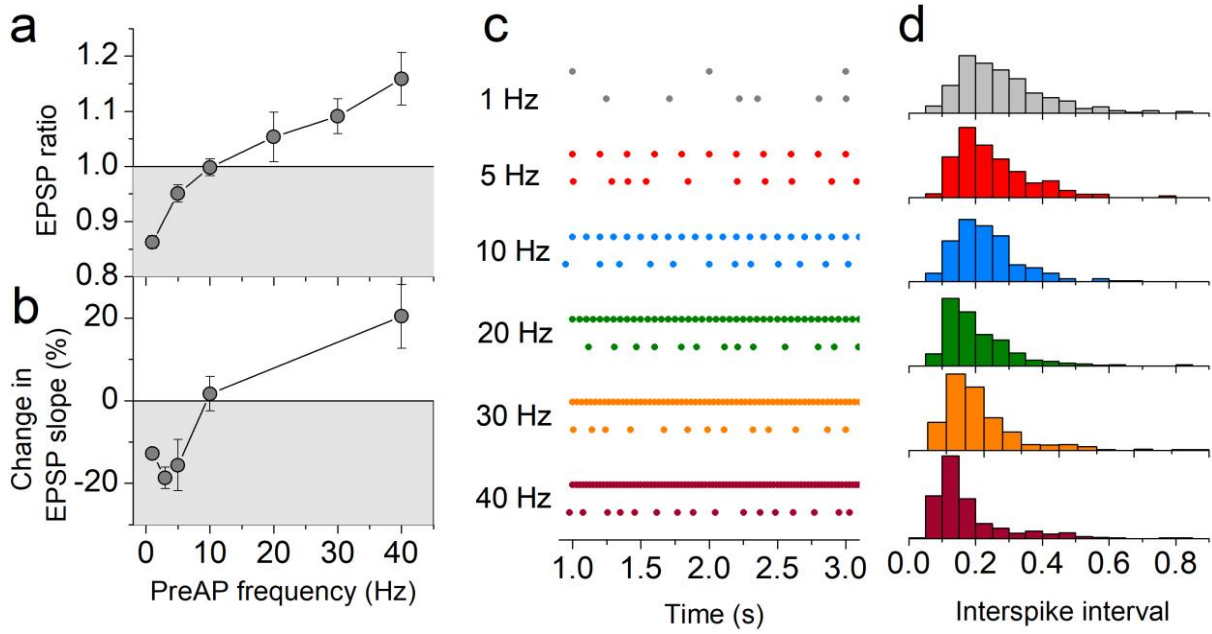

**Figure 3. Long-term plasticity induced by Protocol 2**. (**a**) Long-term plasticity induced by 100 repeatedly applied preAPs at six different rates (1–40 Hz). (**b**) Experimental plasticity behavior of hippocampal neurons induced by Protocol 2. The data are taken from Dudek and Bear (Dudek & Bear, 1992). For each rate in (**a**), the post-neuron fires postAPs to periodic preAPs (the upper plot), which is seen in the lower raster plot in (**c**). (**d**) Irregularity of postAPs, parameterized by the ISI distribution, for the different preAP rates.

To investigate the detailed synaptic plasticity behavior with $\Delta t$, the range of $\Delta t$ was widened (-50 ms – 50 ms), and the results are shown in Figs 2c and d. At frequencies below 10 Hz, LTD and its dependence on the frequency are apparent. Particularly, the crossover between LTD and LTP with $\Delta t$ at 10 Hz (Fig. 2c) reproduces physiological STDP observations(Bi & Poo, 1998; Froemke & Dan, 2002; Markram, et al., 1997; Sjöström, et al., 2001). However, the frequency effect prevails over the temporal order effect in the whole $\Delta t$ range, as shown in Fig. 2d. The experimental data by Sjöström, Turrigiano, and Nelson(Sjöström, et al., 2001) in Fig. 2e indicate the similar tendency; (i) LTD is dominant in the whole $\Delta t$ range at 0.1 Hz, (ii) LTP prevails the LTD irrespective of $\Delta t$ at 40 and 50 Hz, and (iii) an STDP behavior appears at the intermediate frequency of 20 Hz.

## 3.2. Protocol 2

The rate of periodic preAPs in the proposed model was changed from 1 to 40 Hz, altering its contribution to the evolution of the postsynaptic dendritic potential. Alongside the dendritic potential perturbation by such incident periodic preAPs, a stochastic influx of noise current markedly perturbs the dendritic potential in a random manner. The stochastic noise endows the otherwise invariable temporal correlation between preAPs and postAPs with random variability. Nevertheless, it may be that the extent to which the temporal correlation is distorted is dependent on the preAP rate, which is one of the main concerns in Protocol 2.

The simulation results shown in Fig. 3a reveal a homosynaptic LTD when the preAP rate is below approximately 10 Hz, outweighed by LTP at rates larger than 10 Hz. This rate-dependent crossover between LTD and LTP evidently recalls the seminal physiological experiments(Cooper & Bear, 2012; Dudek & Bear, 1992; Philpot, et al., 2007), which are regarded as a direct justification of the BCM theory. The data provided by Dudek and Bear(Dudek & Bear, 1992), re-plotted in Fig.3b, exhibit a transition from LTD to LTP with an increase in preAP frequency, in accordance with the simulation results. It is conceivable that the larger is the preAP rate, the more the preAPs contributes to the dendritic potential evolution over the random noise, establishing a temporal correlation between preAPs and postAPs. Notably, the correlation should capture the causality between preAPs and preAPs in that the preAPs are the direct cause of the postAPs. Thus, the LTD in the low rate regime (<10 Hz) may arise from a weak (or negligible) temporal correlation between the preAPs and postAPs. The two datasets in Figs 3c and d, which plot the timings of preAPs (upper) and postAPs (lower) for each preAP rate (1, 5, 10, 20, 30, and 40 Hz) and the ISI distribution of the postAPs, respectively, underpin this hypothesis. In the low rate regime (<10 Hz), only a weak correlation is established between the preAPs and postAPs due to the disturbance by random noise, which is apparent by the long tail in the ISI distribution compared with those for the higher preAP rates.

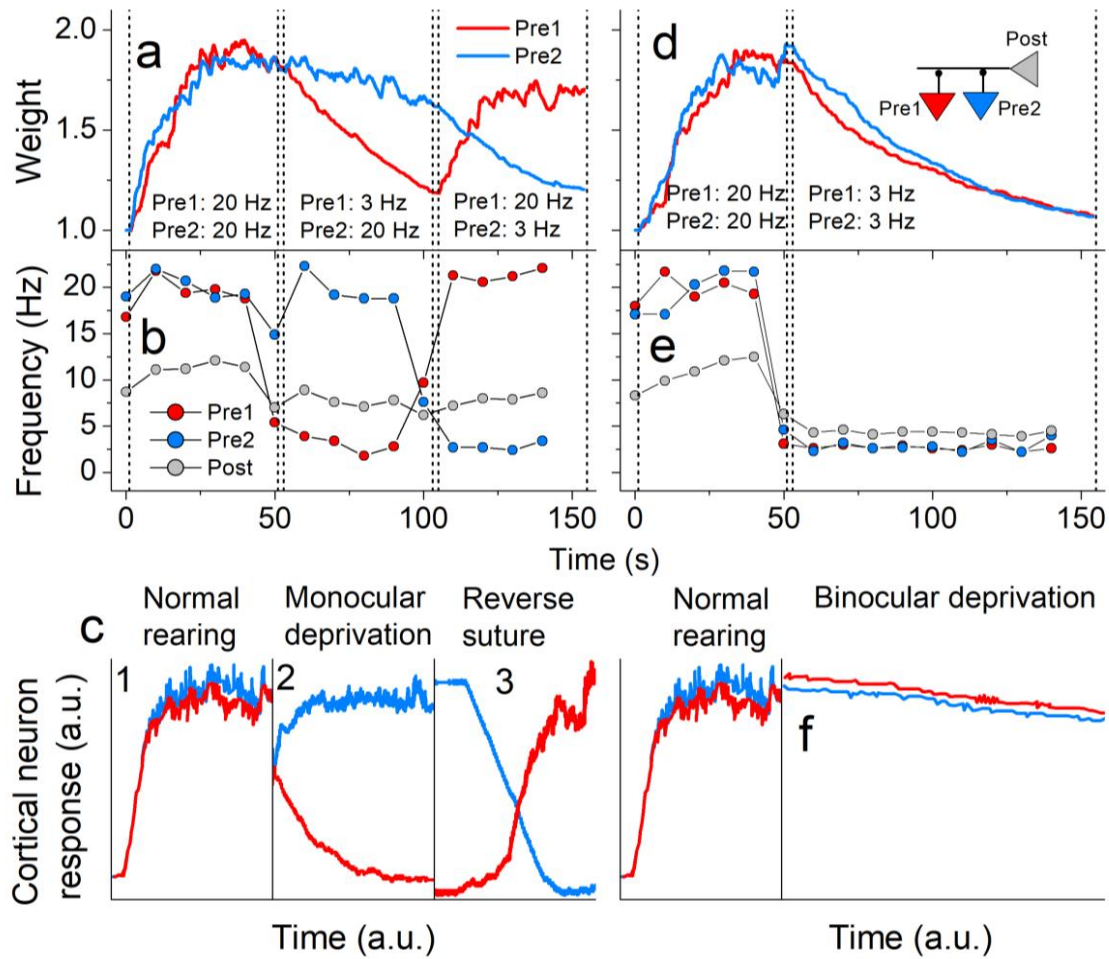

**Figure 4. Long-term plasticity induced by Protocol 3.** Synaptic weight change upon preAPs from two pre-neurons (Pre1 and Pre2) that share the same post-neuron. (**a**) Weight evolution upon Poisson APs from Pre1 and Pre2. The LTP period from the outset for 50 s is followed by a reduction in the AP rate of Pre1 down to 3 Hz while Pre2 maintains 20 Hz. In the subsequent 50 s period, Pre1 and Pre2 are given 20 and 3 Hz rates, respectively. The measured preAP rate for Pre1 and Pre2 and postAP rate are plotted in (**b**). Simulated cortical neuron response during (**c1**) normal rearing, (**c2**) monocular deprivation, and (**c3**) reverse suture using the BCM theory. The data are taken from Cooper and Bear (Cooper & Bear, 2012). (**d**) Weight evolution upon Poisson APs from Pre1 and Pre2 at 20 Hz for 50 s, followed by a 100 s period where both rates are reduced to 3 Hz. The neuronal configuration is sketched in the inset. The monitored preAP and postAP rates are plotted in (**e**). The simulated cortical neuron response during binocular deprivation is plotted in (**f**). The data are taken from Cooper and Bear (Cooper & Bear, 2012).

### 3.3. Protocol 3

Protocol 3 is used to examine the competition between two pre-neurons (denoted by Pre1 and Pre2) that share the same post-neuron. The pre-neurons are given independent preAP rates. To situate them in a more natural condition, the irregularity in preAPs (unlike Protocols 1 and 2) is applied in Protocol 3 such that for each pre-neuron Poisson preAPs are generated at a particular rate. Akin to Protocol 2, the postAPs are partly induced by the incident preAPs. However, the noise current also drives postAP induction. Therefore, their relative contributions to postAP induction are mainly determined by the preAP rate. When applying Protocol 3 to the proposed synaptic plasticity model, we considered two distinct sequences as follows. The first sequence involves three consecutive periods; Pre1 and Pre2 are given a high preAP rate (20 Hz) from the outset (Period 1), the rate of Pre1 drops to 3 Hz while that of Pre2 remains unchanged (Period 2), and the two rates are reversed (Period 3). The second sequence also gives Pre1 and Pre2 a high preAP rate (20 Hz) from the outset, but both rates subsequently drop to 3 Hz.

The simulated behaviors of EPSP ratio change for the two sequences are plotted in Fig. 4. In the first sequence (Fig. 4a), both synapses undergo LTP in first with 20 Hz preAP rates. However, they bifurcate in Period 2 due to the difference in preAP rate: the synapse of Pre1 decreases given that Pre1 has a lower rate than Pre2. The following period (Period 3) indicates the reverse behavior given the reversal of the two preAP rates. Figure 4b displays the monitored preAP rate for Pre1 and Pre2 and the postAP rate. The proposed model leads to competition between the two synapses. Particularly, the rate-dependent competition in Period 2 is a unique feature that distinguishes our model from most of the previous synaptic plasticity models (see more detail in Section 3.5). The key to the competition is the fact that the bAP is *synapse-specific* with regard to the bAP boost. The simulation results match well those of the physiological ocular dominance plasticity behavior, which explains the synaptic weight evolution upon selective monocular deprivation.(Mioche & Singer, 1989) In other words, akin to the eye deprivation of the visual input (by means of eye suture), one of the two pre-neurons is deprived of the 20 Hz preAP rate; instead, it is given the 3 Hz preAP rate as noise. The deprivation leads the synapse to LTD, as shown in Fig. 4a.

Behavior 3 is in excellent agreement with the BCM theory, which successfully accounts for the ocular dominance plasticity behavior. The behavior in Period 1 is equivalent to a normal rearing case, which is also reproduced by the BCM theory as shown in Fig. 4c1. Furthermore, the BCM theory results in a weight change for monocular deprivation and reverse suture, as shown in Figs 4c2 and 3 and similar to Periods 2 and 3 in Fig. 4a, respectively. The results for the BCM theory are taken from Cooper and Bear(Cooper & Bear, 2012).

The simulation results for the second sequence (Fig. 4d) explain the gradual depression of both synapses during the second period, where the preAP rate for both Pre1 and Pre2 is set to 3 Hz (Fig. 4e). This sequence simulates binocular deprivation; both Pre1 and Pre2 are simultaneously deprived of the 20 Hz rate and are given a noise level rate (3 Hz). These results are also consistent with the physiological binocular deprivation plasticity behavior. The binocular deprivation plasticity behavior reproduced by the BCM theory is displayed in Fig. 4f.

## 3.4. Competitive dynamics of [$C$1] and [$C$2] given [$Ca^{2+}$]

The synapses in the proposed model take on either LTP or LTD depending on the amount of calcium influx and the consequent change in [$Ca^{2+}$]. The probability of LTP tends to increase with increasing [$Ca^{2+}$]. LTP and LTD are competitive processes because their mechanisms share the same calcium source, as shown in Fig. 1b. Consequently, the two parallel mechanisms compete for calcium ions while the [$Ca^{2+}$] is limited by the calcium influx through the NMDAR channels. Given the limited [$Ca^{2+}$], the more dominant mechanism retains more calcium ions. In this regard, the key to the LTD-to-LTP transition with an increase in [$Ca^{2+}$] is that $[C1]'$ is a quadratic function of [$Ca^{2+}$], unlike $[C2]'$, which is a linear function of [$Ca^{2+}$] (Eq. (10)). This becomes apparent in the following phase-plane analysis. Note that $[\cdot]'$ denotes the first-order derivative with respect to time.

The phase-plane analysis of [$C$1] and [$C$2] visualizes the dynamics for given [$Ca^{2+}$]. The [$C$1]- and [$C$2]-nullclines are obtained from Eqs (11) and (12), respectively, as follows:

$$[C1] = \tau_{out(C1)} k_{p1}^{f} [Ca^{2+}]^2 \qquad (14)$$

and

$$[C2] = \tau_{out(C2)} k_{d1}^{f}[Ca^{2+}][P]. \tag{15}$$

These nullclines yield the following fixed point

$$([C1],[C2]) = \left(\tau_{out(C1)} k_{p1}^{f}[Ca^{2+}]^2, \tau_{out(C2)} k_{d1}^{f}[Ca^{2+}][P]\right). \tag{16}$$

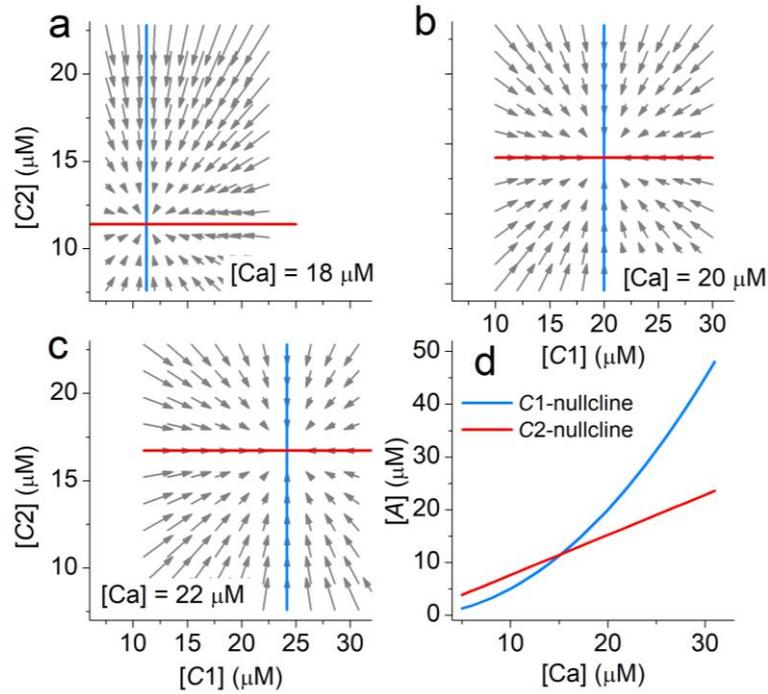

**Figure 5. Phase-plane analysis of [$C1$] and [$C2$] given [$Ca^{2+}$]**. Dynamics of [$C1$] and [$C2$] for [$Ca^{2+}$] of (**a**) 18 μM, (**b**) 20 μM, and (**c**) 22 μM. The blue and red straight lines denote $C1$- and $C2$-nullclines, respectively. In each phase-plane, the intersection point between the $C1$- and $C2$-nullclines is a fixed point. (**d**) Variation in the $C1$- and $C2$-nullclines upon [$Ca^{2+}$].

Three [$C1$]-[$C2$] phase-planes for the three [$Ca^{2+}$] of 18, 20, and 22 μM are shown in Figs 5a, b, and c, respectively. This analysis identifies changes in [$C1$] and [$C2$] (irrespective of their initial values) toward the fixed point for a given [$Ca^{2+}$] value. Thus, in the steady-state, [$C1$] and [$C2$] stabilize at the fixed point. [$C1$] and [$C2$] at the fixed point in Eq. (16) with respect to [$Ca^{2+}$] are plotted in Fig. 5d. Notably, [$C1$] and [$C2$] at the fixed point intersect when

$$[Ca^{2+}] = \frac{\tau_{out(C2)} k_{d1}^f [P]}{\tau_{out(C1)} k_{p1}^f} \tag{17}$$

(Fig. 5d). This means that [$C2$] probably outweighs [$C1$] with [$Ca^{2+}$] below this crossing value [$Ca^{2+}$], whereas the opposite holds for [$Ca^{2+}$] above this crossing value.

Therefore, for a given [$Ca^{2+}$], the steady-state leads [$C1$] and [$C2$] to the fixed point, allowing Eq. (13) to be written as

$$\frac{d[pGluR]}{dt} = \tau_{out(C1)} k_{p1}^f k_{p2}^f ([GluR]_0 - [pGluR])[Ca^{2+}]^2 - \tau_{out(C2)} k_{d1}^f k_{d2}^f [P][pGluR][Ca^{2+}]. \tag{18}$$

Given that LTP is equivalent to $d[pGluR]/dt > 0$, [$Ca^{2+}$]-supporting LTP is derived from Eq. (18) as

$$[Ca^{2+}] > \frac{\tau_{out(C2)} k_{d1}^f k_{d2}^f}{\tau_{out(C1)} k_{p1}^f k_{p2}^f} \cdot \frac{[P][pGluR]}{[GluR]_0 - [pGluR]}. \tag{19}$$

Otherwise, LTD is induced. Therefore, Eq. (19) indicates the threshold [$Ca^{2+}$] for LTP, which varies with the current [pGluR] as opposed to the constant threshold observed with calcium-based plasticity models.(Graupner & Brunel, 2012; Shouval, et al., 2002) The threshold grows with [pGluR] approaching [GluR]$_0$, requiring a larger [$Ca^{2+}$] to induce the same weight change. This is consistent with the BCM theory.(Bienenstock, et al., 1982)

# 4. Discussion

## 4.1. Comparison with previous models

Previous models do not fully reproduce Behaviors 1, 2, and 3, as shown in Table 3. There are various spike-based phenomenological models that consider phenomenological behaviors with various degrees of physiological plausibility.(Clopath, et al., 2010; Froemke & Dan, 2002; Izhikevich & Desai, 2003; Kistler & Hemmen, 2000; Pfister & Gerstner, 2006; Song, et al., 2000) Among them, the simplest models(Froemke & Dan, 2002; Kistler & Hemmen, 2000; Song, et al., 2000) only consider spike-timing data between a preAP and postAP to update synaptic weight in an additive manner. In such STDP models, the weight modification direction is determined by competition between positive (pre-post timing order) and negative (post-pre timing order) contributions.

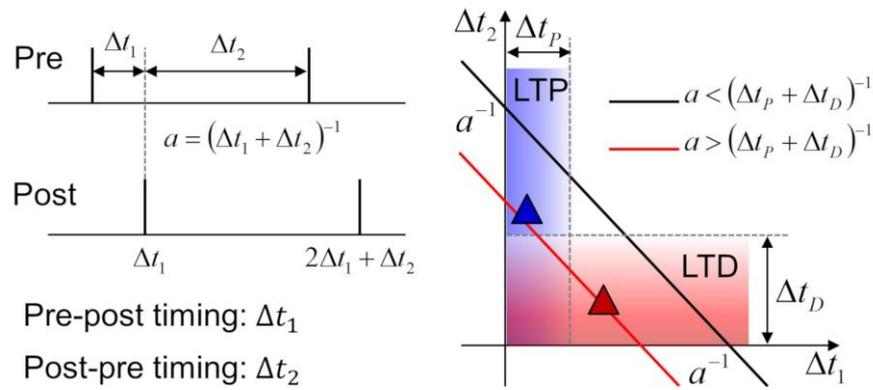

**Figure 6. Pre-post and post-pre timings in Protocol 1 and overlap between the LTP and LTD windows**.

Although they reproduce Behavior 1 well at low frequencies, they fail at high frequencies (rate-dominance) for the following reason. In Protocol 1, when frequency $a$ and pre-post timing $\Delta t_1$ (=$t_{\text{post}} - t_{\text{pre}} > 0$) are independent variables, the post-pre timing $\Delta t_2$ (=$t_{\text{pre}} - t_{\text{post}} > 0$) depends on them:

$$\Delta t_2 = a^{-1} - \Delta t_1. \tag{20}$$

The same holds for the pre-post timing $\Delta t_1$ with $a$ and $\Delta t_2$ taken as independent variables (Fig. 6). Simple STDP models base weight modification on spike-timing entirely. Assuming pre-post timing window $\Delta t_P$ for LTP and $\Delta t_D$ for LTD, LTP and LTD arise when $\Delta t_1 < \Delta t_P$ and $\Delta t_2 < \Delta t_D$, respectively. These are illustrated on a $\Delta t_1$-$\Delta t_2$ plane by the blue and red rectangles in Fig. 6. Also plotted on the same plane for two different frequencies is Eq. (20), where black and red lines designate (i) $a < (\Delta t_P + \Delta t_D)^{-1}$ and (ii) $a > (\Delta t_P + \Delta t_D)^{-1}$, respectively. Case (i) causes no overlap between the LTP and LTD windows irrespective of any ($\Delta t_1$, $\Delta t_2$) pairs on the black line, where the simple STDP models successfully reproduce physiological data. However, the overlap eventually appears with an $a$ of $(\Delta t_P + \Delta t_D)^{-1}$, for instance, 20 Hz with typical $\Delta t_P$ and $\Delta t_D$ of 10 and 40 ms, respectively. The red line shows considerable overlap between the LTP and LTD windows. The models induce exclusive LTP and LTD as far as $\Delta t_2 > \Delta t_D$ (e.g., the blue-filled triangle in Fig. 6) and $\Delta t_1 > \Delta t_P$ (e.g., the red-filled triangle), respectively. Here, there is a major inconsistency with physiological data which indicate exclusive LTP irrespective of timings at such high frequencies.(Sjöström, et al., 2001)

As shown in Fig. 3, at low frequencies of periodic preAPs, the correlation between preAPs and postAPs is weak enough for the random postAPs to prevail over the postAPs. However, with larger frequency, stronger causal correlation established and thus, the induced postAPs dominate the random postAPs. For the postAPs whose association with the preAPs is limited by stochasticity in postAP timing, the probability of the existence of a postAP in a given time bin $\Delta t$ is conceivably modeled as proposed by Izhikevich and Desai(Izhikevich & Desai, 2003), as follows:

$$p(t)\Delta t = [f_s + c(t)]\Delta t, \quad (21)$$

where $f_s$ denotes the rate of Poisson postAPs, which is uniform over time, $c(t)$ indicates an inhomogeneous probability distribution of a postAP due to the preAP-postAP correlation. Here, the preAP timing is taken as $t = 0$. Therefore, $c(t)$ is likely to peak at a particular $t$ (>0) with regard to the temporal preAP-postAP correlation of causality. $c(t)$ tends to increase with the rate of preAPs $f_{pre}$ given that larger $f_{pre}$ induce larger postsynaptic dendritic potentials, which supports the more obvious temporal correlation between the preAP and postAP. For simplicity, we assume $c(t)$ to scale with $f_{pre}$, leading to $c(t) = c_0(t)f_{pre}$. $c_0(t)$ is the probability distribution function of a postAP over time for a $f_{pre}$ of 1 Hz. Thus, Eq. (21) can be re-written as

$$p(t)\Delta t = [f_s + c_0(t)f_{pre}]\Delta t. \tag{22}$$

In the case of all-to-all interaction, the following average change in weight holds(Izhikevich & Desai, 2003)

$$\langle \Delta w \rangle = A_+ \int_0^\infty e^{-t/\tau_+} [f_s + c_0(t)f_{pre}]dt - A_- \int_{-\infty}^0 e^{t/\tau_-} [f_s + c_0(t)f_{pre}]dt, \tag{23}$$

where $A_+$ and $A_-$ are the pre-exponential constants for LTP and LTD, respectively. The weight change decays exponentially with $t$ with time constants $\tau_+$ and $\tau_-$ for LTP and LTD, respectively. Eq. (23) can be rearranged as

$$\langle \Delta w \rangle = (A_+\tau_+ - A_-\tau_-)f_s + f_{pre}\left[A_+ \int_0^\infty e^{-t/\tau_+} c_0(t)dt - A_- \int_{-\infty}^0 e^{t/\tau_-} c_0(t)dt\right]. \tag{24}$$

The first term on the right side of Eq. (24) indicates a weight change without a temporal correlation between preAPs and postAPs. Several physiological observations may show an LTD timing window larger than LTP ($\tau_- > \tau_+$) and $A_-$ comparable to $A_+$. Thus, its contribution to the total weight change may be negative. Regarding the second term on the right side of Eq. (24), the causal temporal correlation between preAPs and postAPs confines nonzero $c_0(t)$ in the region $t > 0$. Therefore, this term may positively contribute to the total weight change, and its contribution eventually outweighs the negative contribution (the first term) with an increase in $f_{pre}$, in agreement with Behavior 2.

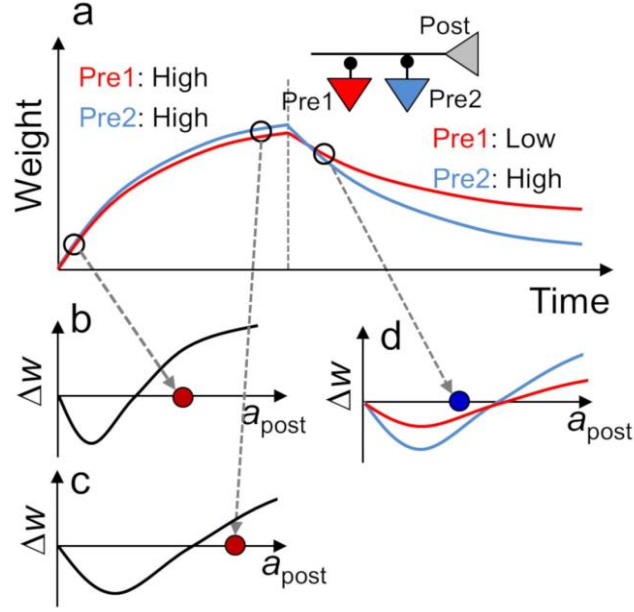

**Figure 7. Schematic of behavior predicted by previous models in response to Protocol 3**. (**a**) Weight evolution upon the firing of two pre-neurons connected to a single post-neuron. Both pre-neurons are given high AP rates in the first period. The consequent modification of the temporary $\Delta w$-$a_{post}$ behavior at two stages (indicated by open circles) is illustrated in (**b**) and (**c**). $\Delta w$ and $a_{post}$ denote a weight change and postAP rate, respectively. The second period in (**a**) displays a large change in weight for Pre2 (high rate) than for Pre1 (low rate) because the $\Delta w$-$a_{post}$ behavior of the models scales with the preAP rate, as illustrated in (**d**).

The STDP model of nearest-neighbor interaction(Izhikevich & Desai, 2003) makes use of the probability of a Poisson postAP at time $t$ for $\Delta t$ as follows:

$$p(t)\Delta t = \left[f_s e^{-|f_s t|} + c_0(t)f_{pre}\right]\Delta t.$$

This differs from Eq. (22) in that a postAP of concern in this model appears in time bin $\Delta t$ (probability = $f_s \Delta t$) and no other postAPs should appear in the range $0 - |t|$ (probability = $e^{-|f_s t|}$). This postAP probability yields an average change in weight of

$$\langle \Delta w \rangle = \left(\frac{A_+ \tau_+}{1+f_s \tau_+} - \frac{A_- \tau_-}{1+f_s \tau_-}\right)f_s + f_{pre}\left[A_+ \int_0^\infty e^{-t/\tau_+} c_0(t)dt - A_- \int_{-\infty}^0 e^{t/\tau_-} c_0(t)dt\right].$$

The first term on the right side indicates the weight change due to Poisson postAPs, which is negative as far as the following condition is satisfied:

$$f_S < \frac{A_-\tau_- - A_+\tau_+}{A_+\tau_+\tau_- - A_-\tau_+\tau_-}.$$

Therefore, these simple plasticity models may account for Behavior 2 and while successful reproduction of Behavior 1 is a prerequisite for Behavior 3, these simple STDP models do not consider Behavior 3.

There are models consistent with the BCM theory regarding the existence of a threshold postAP rate for a transition from LTD to LTP. Commonly, the weight change predicted by these models scales with the preAP rate irrespective of the current threshold value. Additionally, in the models, postAPs (bAPs if applicable) from a post-neuron apply to all relevant pre-neurons in a *synapse-nonspecific* manner. A significant difference between these models and our model is that our model employs the synapse-specific effect of bAPs regarding the bAP boost. In response to monocular deprivation, these two common aspects of the previous models hinder competitive LTD reproduction in Behavior 3 (Fig. 4a, second period). As illustrated in Fig. 7a, the high preAP rates of two pre-neurons in the first period elevate the postAP rate above a threshold for LTP so that the two synapses gain weight in due course. We consider a moving threshold in the illustrations of momentary $\Delta w$-$a_{\text{post}}$ behaviors, as shown in Figs 6b and c, where $\Delta w$ and $a_{\text{post}}$ denote a weight change and postAP rate, respectively. To induce LTD in the models, the postAP rate should decrease below the threshold for LTP, which may be mediated by a reduction in the preAP rate of one of the two pre-neurons (Fig. 7). Because $\Delta w$ for each synapse in these models scales with the preAP rate, a $\Delta w$-$a_{\text{post}}$ behavior for each synapse is distinguishable, as illustrated in Fig. 7d. Notably, in such an LTD condition, the pre-neuron of the higher preAP rate loses more weight than the other (Fig. 7a), which is not consistent with the Behavior 3 in our model (cf. Fig. 4a). This inconsistency highlights the important role of a bAP boost in synaptic competition, mediating different effects of postAPs for the multiple synapses with different dendritic potentials. Although the schematic in Fig. 7 embodies a moving threshold, the same inconsistency holds for the following model with a fixed threshold. The following paragraphs discuss each model in detail.

Izhikevich and Desai proposed a simple condition applied to such simple STDP models in order to successfully account for the BCM theory.(Izhikevich & Desai, 2003) The condition limits the preAP-postAP interaction to the nearest neighbors. Albeit successful in reproducing the BCM theory to some extent, it only holds for stochastic postAPs without a temporal correlation between preAPs and postAPs. Applying high-frequency preAP-postAP pairs of perfect periodicity to this model still produces an STDP behavior as for the aforementioned simple STDP models. Thus, it cannot reproduce Behavior 1. Similar to the simple STDP models, Behavior 2 can be reproduced by this STDP model given the predominant Poisson postAPs at low preAP frequencies (leading to LTD) and a causal temporal correlation between a preAP and postAP at high preAP frequencies (leading to LTP). There exists a threshold postAP-frequency for a transition from LTD to LTP for the nearest neighbor interaction model.(Izhikevich & Desai, 2003) Nevertheless, unlike the BCM theory, the threshold is predetermined by the model parameters.

The aforementioned spike-pair-based STDP model was modified by Pfister and Gerstner to reproduce Behavior 1 (triplet-based STDP model).(Pfister & Gerstner, 2006) They incorporated new pre- and post-neuronal variables that capture the preAP and postAP frequencies, respectively, and that outweigh spike-timing-dependent variables at high frequencies. Therefore, these new variables underlie the reproduction of Behavior 1 as well as the BCM theory. However, this model is given a fixed threshold for a transition from LTD to LTP. Behavior 2 appears to be reserved for this mode because of its consistency with the BCM theory. In other words, high-frequency preAPs induce high-frequency postAPs above the threshold for LTP whereas low-frequency preAPs are unlikely to allow the postAP frequency to exceed the threshold, leading to LTD. Despite the similarity to the BCM theory, this model was modified to incorporate a moving (rather than fixed) threshold for LTP, taking a step toward the BCM theory.(Gjorgjieva, et al., 2011)

The voltage-based STDP model(Clopath, et al., 2010) associates synaptic plasticity with post-neuronal somatic potential. Note that the neurons in the model are regarded as point neurons, and thus no dendritic potential is considered. The potential-related variables enable the rate-dependent plasticity behavior to dominate the timing-dependent behavior at high preAP and postAP frequencies, reproducing Behavior 1 and the BCM theory. Notably, the threshold postAP frequency for LTP in this model varies with the post-neuronal membrane potential averaged over the past second such that the LTD factor is enhanced by the average membrane potential. The successful realization of the BCM theory is related to Behavior 2 in this model.

Calcium-based models consider the calcium concentration the direct cause of synaptic plasticity.(Graupner & Brunel, 2012; Shouval, et al., 2002) A threshold calcium concentration ($[Ca^{2+}]_{th}$) for LTD and LTP is fixed to a certain level; a general strategy endows LTP with a $[Ca^{2+}]_{th}$ higher than that for LTD based on physiological observations.(Cho, et al., 2001; Hansel, et al., 1997; J Lisman, 1989) Each model in this category differs in the detail of the mechanism used for encoding spike-timing and AP rate information as $[Ca^{2+}]$. The unified model by Shouval, Bear, and Cooper(Shouval, et al., 2002) is the first to consider the crucial and long-lasting (a few tens of milliseconds) contribution of bAPs to the postsynaptic dendritic potential, which consequently dictates the calcium influx through NMDAR channels. However, unlike in our model, bAP boosts by the dendritic potential are ignored in this model. Furthermore, it considers an unrealistically negligible and weight-independent contribution of preAPs to the dendritic potential.

This model reproduces Behaviors 1 and 2, increasing its capability of realizing the BCM theory. A threshold postAP frequency for LTP varies with the current weight, in that larger current weight requires higher postAP frequency to induce a further increase in the weight. Similar to all previous models, a post-neuron in connection to multiple pre-neurons provides synapses with synapse-nonspecific signals; the bAPs from a single somatic AP firing event have the same amplitude and width irrespective of the current synaptic weight (or equivalent variable reflecting the current weight).

Graupner and Brunel proposed a simpler model of synaptic plasticity following the seminal model by Shouval, Bear, and Cooper.(Graupner & Brunel, 2012) This model also uses fixed $[Ca^{2+}]_{th}$ values for LTP and LTD ($[Ca^{2+}]_{th}$ for LTP > $[Ca^{2+}]_{th}$ for LTD) and transient $[Ca^{2+}]$ (decaying out) independently induced by preAPs and postAPs. The total $[Ca^{2+}]$ at a given time is simply the linear sum of all $[Ca^{2+}]$ trajectories evoked by past preAPs and postAPs. LTP in this model is not the consequence of coincidence detection because preAPs (postAPs) can only evoke LTP in the absence of their counter APs as far as the induced $[Ca^{2+}]$ exceeds the threshold for LTP. Behaviors 1 and 2 and the BCM theory were proven to be reproduced by this model. Furthermore, this model considers a moving threshold for LTP given the current weight-dependent relative contributions of LTP and LTD to the total weight.

## 4.2. Model as an embedded learning algorithm in neuromorphic hardware

Practically, synaptic plasticity models are of significant concern in neuromorphic hardware with embedded learning ability.(Kornijcuk & Jeong, 2019; Neftci, 2018) Embedded learning requires learning implementation using dedicated hardware (neuromorphic processors) without general-purpose computers. The key to neuromorphic processors is the *parallel* computing of neural processing and synaptic plasticity over *parallel* cores as opposed to von Neumann architecture highlighting *serial* computing.(Davies, et al., 2018; Furber, Galluppi, Temple, & Plana, 2014; Indiveri & Liu, 2015; Kornijcuk & Jeong, 2019; S. Moradi, 2018) Although each core has a von Neumann architecture, the lack of memory- and processor-sharing among parallel cores ensures parallel computing, which avoids the notorious von Neumann bottleneck. However, this requirement imposes severe constraints on embedded learning algorithms including (i) locality, (ii) event-driven update, and (iii) minimal use of variables. The locality ensures the use of topological and temporal local variables to update weights independently from variables allocated in other cores. The update upon events (presynaptic and/or postsynaptic spikes) leverages the parallel architecture in neuromorphic hardware by simultaneous updates in the cores, in contrast to backpropagation algorithms for layer-wise update. The last constraint is the limited memory capacity of each core. The more variables an algorithm considers, the larger is the memory required per core. Further, the memory demand increases super-linearly with the number of variables because several variables should be duplicated over cores.(Kornijcuk & Jeong, 2019)

A considerable number of synaptic plasticity rules (all rules addressed in this study) are event-driven algorithms of locality, meeting the first two constraints. This is one of the key reasons for the extensive focus given to synaptic plasticity rules. For instance, STDP models have been frequent subjects of neuromorphic hardware implementations using various designs.(Azghadi, Iannella, Al-Sarawi, Indiveri, & Abbott, 2014; Davies, et al., 2018; Kornijcuk, et al., 2017; Kornijcuk, et al., 2019; Lammie, Hamilton, Schaik, & Azghadi, 2019; Pedroni, et al., 2019) Particularly, the pair-based STDP model(Song, et al., 2000) and triplet models(Pfister & Gerstner, 2006) base weight updates on two variables (one pre- and one post-synaptic) and four variables (two pre- and two post-synaptic variables), respectively, so that the STDP models satisfy the three aforementioned constraints. Nevertheless, apart from the constraints on an embedded learning algorithm, the algorithm should be able to train the SNN according to the purpose. However, as addressed in Section 3.5, the inconsistency between plasticity behaviors with reference to physiological observations is obvious for the STDP models, making them unsuitable as general learning algorithms valid under diverse spiking conditions.

The proposed simplified calcium signaling cascade model offers a unified learning framework that is consistent with the physiological observations under various spiking conditions. The locality is perfectly satisfied because the weight update is evaluated using eight topological and temporal local variables ($u_m^d$, $P_{\text{open}}$, $s_i^d$, $s_i^b$, [$Ca^{2+}$], [$C1$], [$C2$], and [pGluR] on a current time step). Note that [GluR] is not an independent variable because [GluR] = [GluR]$_0$ – [pGluR]. Additionally, the weight update in the proposed model is triggered by pre and postsynaptic events (spikes). These attributes highlight the proposed model as an appropriate learning algorithm for neuromorphic hardware. However, the eight independent variables incorporated into the proposed model cause a larger memory overhead in a neuromorphic processor core than with other algorithms (the number of variables in different models is listed in Table 4). This motivates a further simplification and abstraction of the proposed model while maintaining the reproducibility of physiological observations. For the moment, we leave the task as a future work because it is beyond the scope of the present study. However, the proposed model serves as a steppingstone to a memory-efficient learning algorithm with fidelity to physiological synaptic plasticity behaviors.

### 4.2. Knowledge learning and discovery with the proposed learning rule

Generally, knowledge is stored as synaptic weights in a neural network. Event-driven learning rules, such as the proposed rule in this study, enable new knowledge to be learned in addition to pre-existing knowledge in the network, referred to as incremental learning. Because the knowledge-learning capacity is limited by the limited numbers of neurons and synapses, learning new knowledge may destroy pre-existing knowledge undesirably. One extreme way to avoid the disturbance is to use an exceedingly large network as for hierarchical temporal memory (HTM) networks(Hawkins & Ahmad, 2016). A large amount of neurons and synapses in the HTM network significantly reduces the probability of an overlap between different knowledge. However, considering the limited network size in practice, this extreme method may be prone to learning new knowledge. To leverage the knowledge capacity for a given network size, dense (rather than sparse) representation of knowledge is desirable, which causes overlaps between different knowledge inevitably. Appropriate learning rules let the different knowledge compete over the limited resources when overlapping, thereby balancing the representations of different knowledge depending on the causality between pre- and postsynaptic events. In this regard, the proposed model may be an appropriate learning rule because of its competition-based plasticity behavior based on the causality as shown in Behavior 3.

The key implication in all three plasticity behaviors (Behaviors 1, 2, and 3) is that the synaptic weight learns such that the causality between pre- and postsynaptic events strengthens the weight while the acausality weakens the weight. The proposed model finds the causality defined in both time and activity domains consistently, which is the most distinct feature from the previous models as discussed in Section 4.1. Although the learning examples in Behaviors 1, 2, and 3 are categorized as unsupervised learning, the proposed model can also be applied to supervised learning by injecting a supervision current signal into a postsynaptic neuron, which evokes a spike(s) from the postsynaptic neuron. Given the causality captured by the weight, supervised learning (for classification) with a single layer SNN maps input neurons (features) onto the output neurons (classes), indicating the causality between each feature and classes. This may reveal key features to each class, i.e., knowledge discovery. Note that input features encoded using a temporal code or activity code(Jeong, 2018) because the proposed model captures the causality in both domains. However, such direct causality discoveries appear impossible for multilayer SNN, where the input features are connected to the output classes indirectly through hidden neurons whose meanings are unknown. In this regard, the evolving connectionist system (ECOS) framework(Nikola Kasabov, 1998) may offer a chance to discover the rules underlying complex data when evolving fuzzy neural network is trained using the proposed model. We leave this task as a future work for the moment.

## 5. Conclusions

We proposed a synaptic plasticity model capable of reproducing plasticity behaviors in response to (i) the standard STDP protocol at various rates of preAP-postAP pairs, (ii) periodic preAPs at various rates, and (iii) monocular and binocular deprivation. The model incorporates a calcium signaling cascade in an efficient manner that significantly simplifies the full pathways by considering only a few essential attributes. Based on the competition between phosphorylation and dephosphorylation of glutamate receptors given the momentaneous calcium concentration, we realized a history-dependent threshold of the calcium concentration for transition from homosynaptic LTD to LTP. This distinguishes our model from the previous calcium-based plasticity models, which consider a fixed calcium concentration threshold for LTP and LTD. Another major feature of our model is the implementation of a bAP boost underlying the synapse-specific effect of postAPs when multiple pre-neurons (and thus multiple synapses) share a single post-neuron, which is the key to competitive LTD between such multiple synapses in line with monocular deprivation experiments. This distinguishes our model from not only other calcium-based models but also spike-based phenomenological models. Lastly, we carefully hypothesize one of the functional implications of bAP boosts in physiology from the present study, namely a means of synaptic competition and the implementation is simplified in our model. Underpinning this hypothesis may require modeling tbAP behavior in detail with special attention to bAP attenuation and delay along the dendrite and the spatial distribution of dendritic spines.

**Funding**: This work was supported by the National Research Foundation of Korea (grant number NRF-2018K2A9A2A08000151)

**The authors declare no conflict of interest.**

**Table 1.** Model parameters

| Symbol | Value | Symbol | Value | Symbol | Value |
|---|---|---|---|---|---|
| Neuronal parameters (soma) | | | | | |
| $C_m^s$ | 80 mF | $g_l^s$ | 2 S | $g_{e0}^s$ | 0.2 S |
| $s_{i0}^s$ | 1.2 | $\tau_s$ | 4 ms | $\Delta g_{adp}^s$ | 0.5 S |
| $\tau_{adp}$ | 100 ms | $u_{th}$ | -59 mV | $t_{ref}$ | 4 ms |
| $u_l$ | -65 mV | $u_{rest}$ | -65 mV | $u_{adp}$ | -70 mV |
| $\mu_i$ | 9 mA | $\sigma_i$ | 72 mA | | |
| Neuronal parameters (dendrite) | | | | | |
| $C_m^d$ | 40 mF | $g_l^d$ | 2 S | $g_{e0}^d$ | 0.2 S |
| $s_{i0}^d$ | 4.2 | $\tau_d$ | 2 ms | $\tau_{bp}$ | 2 ms |
| $s_0^b$ | 90 | $u_{th}^b$ mean | -58 mV | $u_{th}^b$ s.d. | 5.8 mV |
| $\mu_i$ | 9 mA | $\sigma_i$ | 18 mA | $s_1^b$ | 10 |
| Calcium influx | | | | | |
| $t_{open}$ | 100 ms | $G_{Ca0}$ | 70 mS | $k_{Ca}$ | 250 V$^{-1}$ |
| $u_{Ca}$ | -40 mV | $\tau_{out(Ca)}$ | 200 ms | | |
| C1 evolution | | | | | |
| $k_{p1}^f$ | 2.5×10$^5$ M$^{-1}$s$^{-1}$ | $\tau_{out(C1)}$ | 200 ms | $[C1]_0$ | 341.9 nM |
| C2 evolution | | | | | |
| $k_{d1}^f$ | 1.9×10$^6$ M$^{-1}$s$^{-1}$ | $[P]$ | 2 μM | $\tau_{out(C2)}$ | 200 ms |
| $[C2]_0$ | 340 nM | | | | |
| Phosphorylation of GluR | | | | | |
| $k_{p2}^f$ | 70×10$^2$ M$^{-1}$s$^{-1}$ | $[GluR]_0$ | 10 μM | $k_{d2}^f$ | 20×10$^3$ M$^{-1}$s$^{-1}$ |
| $[pGluR]_0$ | 2 μM | | | | |

**Table 2.** Summary of protocols

|  | Neuronal topology | Variables | |
|---|---|---|---|
|  |  | preAP | bAP |
| Protocol 1 | 1pre-1post | Timing and rate (with perfect periodicity) | Timing and rate (with perfect periodicity) |
| Protocol 2 | 1pre-1post | Rate (with perfect periodicity) | - (induced bAP) |
| Protocol 3 | 2pre-1post | Difference in mean firing rate between the two pre-neurons (Poisson AP) | - (induced bAP) |

**Table 3**. Comparison with previous models

| Model | Protocol 1 | Protocol 2 | Protocol 3 | Threshold for LTP |
|---|---|---|---|---|
| Spike-based phenomenological models | | | | |
| Spike pair-based model (all-to-all interaction)(Froemke & Dan, 2002; Kistler & Hemmen, 2000; Song, et al., 2000) | No | Yes | No | - |
| Izhikevich model (nearest-neighbor interaction)(Izhikevich & Desai, 2003) | No | Yes | No | Fixed |
| Triplet models(Gjorgjieva, et al., 2011; Pfister & Gerstner, 2006) | Yes | Yes | No | Fixed(Pfister & Gerstner, 2006) Moving (Gjorgjieva, et al., 2011) |
| Voltage-based STDP(Clopath, et al., 2010) | Yes | Yes | No | Moving |
| Calcium-based models | | | | |
| Shouval et al.(Shouval, et al., 2002) | Yes | Yes | No | Moving |
| Graupner and Brunel(Graupner & Brunel, 2012) | Yes | Yes | No | Moving |
| Proposed model | Yes | Yes | Yes | Moving |

**Table 4**. Number of variables required for various models

| Model | Number of variables | Locality of variables |
|---|---|---|
| Spike-based phenomenological models | | |
| Spike pair-based model (Froemke & Dan, 2002) | 4 (two pre- and two post-synaptic variables) | All local |
| Spike pair-based model (Song, et al., 2000) | 2 (one pre- and one post-synaptic variable) | All local |
| Izhikevich model (Izhikevich & Desai, 2003) | 2 (one pre- and one post-synaptic variable) | All local |
| Triplet models (Gjorgjieva, et al., 2011) | 4 (one pre- and two post-synaptic variables; history-dependent coefficient of LTD) | 3 local 1 temporal global (history-dependent coefficient of LTD) |
| Triplet models (Pfister & Gerstner, 2006) | 4 (two pre- and two post-synaptic variables) | All local |
| Voltage-based STDP (Clopath, et al., 2010) | 5 (membrane potential; two low pass-filtered potentials; low pass-filtered spike train; history-dependent coefficient of LTD) | 4 local 1 temporal global (history-dependent coefficient of LTD) |
| Calcium-based models | | |
| Shouval et al. (Shouval, et al., 2002) | 7 ($[Ca^{2+}]$; two exponential functions of time for $i_{ca}$ evaluation; four exponential functions of time for membrane potential evaluation) | All local |
| Graupner and Brunel (Graupner & Brunel, 2012) | 1 ($[Ca^{2+}]$) | All local |
| Proposed model | 8 ($u_m^d$, $P_{open}$, $s_i^d$, $s_i^b$, $[Ca^{2+}]$, $[C1]$, $[C2]$, and $[pGluR]$) | All Local |